# Rectangular finite elements for modeling the mechanical behavior of auxetic materials

Alexey V. Mazaev

alexeymazaev@outlook.com

**Abstract**

This paper is devoted to the exploration of rectangular finite elements' ability to model the stress-strain state of isotropic and orthotropic materials with a negative Poisson's ratio, known as auxetic materials. By employing linear elasticity in the plane stress formulation, the research evaluates the linear compatible and the quadratic incompatible shape functions in describing the mechanical behavior of auxetic materials within a displacement-based finite element method under static shear and indentation. Additionally, the analytical expression of an incompatible rectangular finite element is adapted to accommodate an orthotropic case. Hexachiral and re-entrant honeycomb structures, characterized by auxetic behavior, are modeled as continuous media with homogenized properties using analytical expressions for their effective material constants. The findings reveal that while the classical shape functions may be sufficient for displacement modeling, they are ineffective in accurately predicting the characteristic auxetic behavior and stress distributions in auxetic materials. In contrast, the incompatible shape functions prove to be effective in providing appropriate stress modeling in both cases. This work underscores the relevance of the incompatible rectangular finite elements in the analysis of advanced materials with a negative Poisson's ratio. It provides computationally efficient approaches for the calculation of auxetic honeycomb structures and multilayer composites based on them.

## 1. Introduction

Auxetic materials have been gaining popularity since the first experimental observation around four decades ago and continue to attract the attention of researchers nowadays [1-8]. This long-lasting interest is driven by their unique mechanical properties for broad potential applications across traditional industries, including, but not limited to, aerospace, automotive, civil engineering, and healthcare. Auxetic materials exhibit a negative Poisson's ratio in contrast to ordinary materials, which leads to inverse lateral behavior under tensile loads. This in turn provides improved mechanical properties in shear resistance, toughness, energy absorption, damping, and more [1-8]. The studies summarized below focus on innovative designs and computational approaches that enrich the understanding of synthetic auxetic materials, in particular honeycomb structures, which are viable as effective engineering solutions.

Bezazi et al. [9] introduced a center-symmetric honeycomb configuration that accounts for manufacturing constraints typical of production methods like resin transfer molding and rapid prototyping. By adjusting the base wall ratio and angle, they demonstrated substantial changes in in-plane stiffness and Poisson's ratio. The inclusion of edge corners in the unit cell resulted in increased flexibility compared to classical configurations. Their analytical model showed good agreement with the Gibson and Ashby rib-bending model when manufacturing constraints were considered, suggesting the potential of their design for multifunctional honeycomb cores in sandwich components.

Grima et al. [10] proposed a mechanism for achieving auxetic behavior in foam materials through the rotation of rigid units. This mechanism differs from previous models as it does not require structural modifications at the joints or rib breakage. Scanning electron microscope images and finite element modeling supported the feasibility of this mechanism. They acknowledged that real foams might exhibit more complex behaviors but posited that the rotation of rigid units is a predominant mechanism responsible for the auxetic effect in foams.

Lee et al. [11] investigated rotational particle structures exhibiting negative Poisson's ratio (NPR) using finite element analysis. By varying design variables such as the ratio of fibril length to particle diameter and the fibril angle, they demonstrated that the Poisson's ratio could be effectively controlled. Their study provided insights into the relationship between geometric parameters and the mechanical properties of rotational particle structures, offering guidelines for material design. They also verified that synthesized polymeric materials could possess particle-fibril molecular structures similar to their finite element models.

Alderson et al. [12, 13] focused on the elastic constants of chiral and anti-chiral honeycombs. They developed finite element models and manufactured honeycombs using rapid prototyping to characterize their mechanical properties. Their studies identified deformation mechanisms involving cylinder rotation and ligament flexing, contributing to auxetic behavior. They showed that honeycombs with higher ligament coordination numbers exhibited increased Young's moduli and that certain configurations could achieve Poisson's ratios close to -1. The research also introduced re-entrant cylinder-ligament honeycombs displaying negative Poisson's ratios and synclastic curvature upon out-of-plane bending.

Wang et al. [14] analyzed the microstructural effects on the Poisson's ratio of star-shaped two-dimensional systems. Their finite element analysis showed that re-entrant cell shapes alone do not guarantee a negative Poisson's ratio. Instead, auxeticity depended on the auxetic angle, with angles greater than about 20 degrees necessary to achieve NPR. Their work emphasized the importance of precise geometric control in designing auxetic materials, noting that larger auxetic angles increased auxeticity but reduced the effective shear modulus.

Jiang et al. [15] investigated the limiting strain for auxeticity under large compressive deformation in chiral and re-entrant cellular solids. Through experiments and simulations, they found that re-entrant honeycombs could only preserve auxetic effects under small compressive strains (<3-4%) due to instability and self-contact between ribs. In contrast, chiral cellular solids maintained auxetic behavior under much larger compressive strains (up to 10-30%), attributed to chirality-induced rotation. These findings provide useful guidelines for designing auxetic materials that sustain auxetic effects under larger deformations.

Tan et al. [16] combined auxetic structures with hierarchical honeycombs to enhance crashworthiness. Their re-entrant hierarchical honeycombs exhibited significantly improved specific energy absorption compared to classic re-entrant honeycombs under in-plane impact. Their parametric studies showed that the energy absorption performance depended on impact velocity and relative density, providing valuable insights for designing lightweight, high-performance structures. The proposed designs improved mean crushing force and maintained negative Poisson's ratio.

Koutsianitis et al. [17] explored the creation of band gaps in conventional and star-shaped auxetic materials. Using finite element analysis and optimization algorithms, they demonstrated that specific microstructures could lead to the occurrence of band gaps. Their findings indicated that both conventional and auxetic microstructures could exhibit band gaps, with shape optimization playing a crucial role in achieving desired frequency isolation. The study suggests that optimization algorithms like genetic algorithms are effective in selecting design parameters for targeted frequency ranges.

Dhari et al. [18, 19] studied the deformation mechanisms of re-entrant honeycomb auxetics under inclined loading. They identified new transitional deformation stages and micro-modes that affected the mechanical response. Their plastic hinge tracing method provided a systematic way to analyze deformation, revealing that inclined loading induced localized plasticity and altered energy dissipation characteristics. The research is particularly relevant for designing structures subjected to non-symmetrical or dynamic loads, as it highlights how inclined loads induce new response mechanisms in auxetic structures.

Rezaei et al. [20] used the modified Solid Isotropic Material with Penalization (SIMP) topology optimization method to design diverse auxetic metamaterials with hyperelastic properties. Their results showed that the geometry of auxetic structures heavily depended on optimization parameters, including initial designs and volume fractions. By producing and testing structures with symmetric unit cells, they validated their designs and highlighted potential applications ranging from biosensors to phononic crystals. The deformation mechanisms were based on either bending or buckling, depending on the structure.

Nascimento et al. [21] applied the Radial Point Interpolation Method (RPIM) to simulate the elasto-static behavior of honeycomb structures. Comparing RPIM with the finite element method, they found that RPIM could provide more accurate approximations for complex geometries, despite slightly higher computational costs. Their work underscores the importance of advanced numerical methods in analyzing intricate structural architectures and contributes to the state-of-the-art concerning meshless methods.

Luo et al. [22] introduced a negative Poisson's ratio lattice structure combining chiral and re-entrant properties, inspired by ancient Chinese window grills. Through experiments and simulations, they demonstrated a significant NPR effect dependent on geometric parameters. They observed that the strongest NPR effect occurred at specific length ratios and deflection angles, with the Poisson's ratio reaching -0.6. Their design approach offers a framework for developing new types of architected metamaterials with potential applications in various fields, including construction engineering and biomedical sensors.

Zawistowski and Poteralski [23] performed parametric optimization of selected auxetic structures to identify geometrical parameters influencing the Poisson's ratio. Using numerical simulations and the Multi-Objective Genetic Algorithm (MOGA), they increased the auxetic effect and reduced equivalent stresses. This approach enables the manipulation of material properties by adjusting geometric parameters.

In scientific papers [9-23], mechanical simulations have applied the finite element method, specifically employing plane elements to directly describe the geometrical shapes of auxetic honeycombs and structures. In addition, a widely adopted approach uses three-dimensional finite elements to model the in-plane mechanical behavior of auxetic honeycombs and their derived composites through commercial software [24-32]. However, an active area of research, often conducted in parallel, focuses on deriving effective material constants for various honeycomb configurations via homogenization techniques [12, 13, 24, 33-40] aimed at simplified and computationally efficient simulations.

This paper investigates the application of rectangular finite elements to model the stress-strain state of isotropic and orthotropic materials with a negative Poisson's ratio within the plane stress formulation of linear elasticity under static shear and indentation. The study employs the classical compatible shape functions for linear interpolation of field variables and the incompatible shape functions for quadratic interpolation in the displacement-based finite element approach. Along with this, hexachiral and re-entrant honeycomb structures, known for their negative Poisson's ratio, are modeled as continuous media with homogenized properties by analytical expressions for determining their effective material constants. The results from the finite element modeling are presented via graphs and contour plots, which are subsequently analyzed and discussed.

## 2. Derivation of rectangular finite elements for isotropic and orthotropic materials

Consider a rectangular finite element with four nodes located at its vertices, as illustrated in Figure 1. Let $a_{fe}$ and $b_{fe}$ denote the side dimensions of the finite element along the $x$- and $y$-axes, respectively. The coordinates of its center are $x_c$ and $y_c$. The displacement components of the nodes along the $x$- and $y$-axes are denoted by $v^x$ and $v^y$, respectively.

The dimensionless coordinates $\xi$ and $\eta$ of the rectangular element are defined as: $\xi = 2(x - x_c)/a_{fe}$, and $\eta = 2(y - y_c)/b_{fe}$. Thus, the coordinates of the four nodes are as follows: $\xi_1 = -1$, $\eta_1 = -1$, $\xi_2 = 1$, $\eta_2 = -1$, $\xi_3 = 1$, $\eta_3 = 1$, and $\xi_4 = -1$, $\eta_4 = 1$.

The analytical form of the finite element stiffness matrix $k^e$ is derived through integration, as shown in equation [41-44]

$$k^e_{r,s} = \int \beta_r^T \chi \beta_s h \, dx \, dy = \frac{a_{fe} b_{fe} h}{4} \int_{-1}^{1} \int_{-1}^{1} \beta_r^T \chi \beta_s \, d\xi \, d\eta \,. \tag{1}$$

Here, $r$ and $s$ are indices of the matrix blocks $(r, s = 1, 2, 3, 4)$, $h$ is the out-of-plane size of the finite element, and $\beta_r$ (and similarly $\beta_s$) is the matrix relating nodal displacements to strains

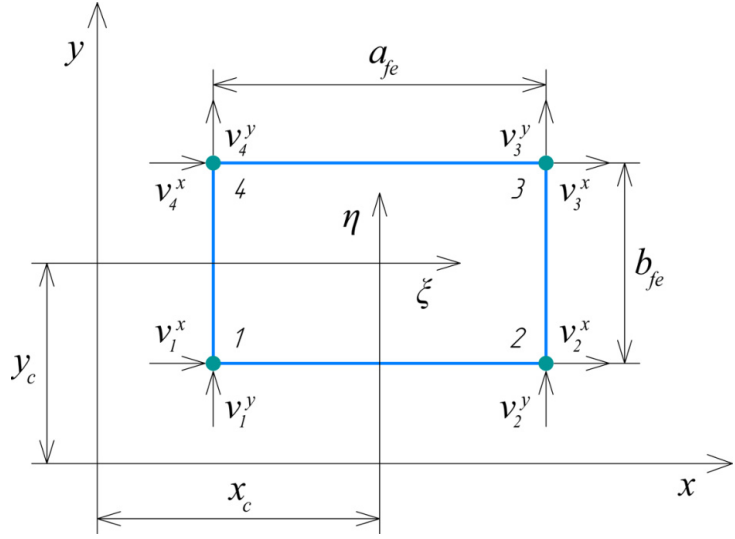

**Fig. 1.** Parameters of a rectangular finite element [45].

$$\beta_r = L\alpha_r = \frac{1}{2}\begin{pmatrix} \dfrac{\xi_r(1+\eta_r\eta)}{a_{fe}} & 0 \\ 0 & \dfrac{\eta_r(1+\xi_r\xi)}{b_{fe}} \\ \dfrac{\eta_r(1+\xi_r\xi)}{b_{fe}} & \dfrac{\xi_r(1+\eta_r\eta)}{a_{fe}} \end{pmatrix}. \tag{2}$$

$L$ is the matrix differential operator

$$L = \begin{pmatrix} \partial/\partial x & 0 \\ 0 & \partial/\partial y \\ \partial/\partial y & \partial/\partial x \end{pmatrix}, \tag{3}$$

and $\alpha_r$ is the matrix of classical compatible shape functions using linear interpolation

$$\alpha_r = \begin{pmatrix} \dfrac{(1+\xi_r\xi)(1+\eta_r\eta)}{4} & 0 \\ 0 & \dfrac{(1+\xi_r\xi)(1+\eta_r\eta)}{4} \end{pmatrix}. \tag{4}$$

$\chi$ is the elasticity matrix for an isotropic material under plane stress

$$\chi = \frac{E}{1-\mu^2}\begin{pmatrix} 1 & \mu & 0 \\ \mu & 1 & 0 \\ 0 & 0 & \dfrac{1-\mu}{2} \end{pmatrix}, \tag{5}$$

where $E$ is elastic modulus and $\mu$ is Poisson's ratio.

It is convenient to split the stiffness matrix into two terms

$$k_{r,s}^e = k_{r,s}^E + k_{r,s}^G, \tag{6}$$

where $k^E$ denotes the submatrix of normal deformations

$$k_{r,s}^E = \frac{a_{fe} b_{fe} h}{4} \int_{-1}^{1} \int_{-1}^{1} \beta_r^T \chi_E \beta_s \, d\xi \, d\eta, \tag{7}$$

and $k^G$ denotes the submatrix of shear deformations

$$k_{r,s}^G = \frac{a_{fe} b_{fe} h}{4} \int_{-1}^{1} \int_{-1}^{1} \beta_r^T \chi_G \beta_s \, d\xi \, d\eta, \tag{8}$$

$\chi_E$ and $\chi_G$ are the matrices of elastic coefficients for an isotropic material, defined by

$$\chi_E = \frac{E}{1-\mu^2} \begin{pmatrix} 1 & \mu & 0 \\ \mu & 1 & 0 \\ 0 & 0 & 0 \end{pmatrix}, \tag{9}$$

and

$$\chi_G = G \begin{pmatrix} 0 & 0 & 0 \\ 0 & 0 & 0 \\ 0 & 0 & 1 \end{pmatrix}, \tag{10}$$

$G$ is shear modulus, defined by $G = E/2(1+\mu)$.

Calculating the integrals in equations (7) and (8) yields, respectively

$$k_{r,s}^E = \frac{Eh}{4(1-\mu^2)} \begin{pmatrix} \gamma \xi_r \xi_s \left(1 + \frac{\eta_r \eta_s}{3}\right) & \mu \xi_r \eta_s \\ \mu \eta_r \xi_s & \frac{\eta_r \eta_s}{\gamma} \left(1 + \frac{\xi_r \xi_s}{3}\right) \end{pmatrix}, \tag{11}$$

$$k_{r,s}^G = \frac{Gh}{4} \begin{pmatrix} \frac{\eta_r \eta_s}{\gamma} \left(1 + \frac{\xi_r \xi_s}{3}\right) & \eta_r \xi_s \\ \xi_r \eta_s & \gamma \xi_r \xi_s \left(1 + \frac{\eta_r \eta_s}{3}\right) \end{pmatrix}. \tag{12}$$

The sum of these provides the analytical expression for the stiffness matrix of the classical compatible finite element for an isotropic material, where $\gamma = b_{fe}/a_{fe}$ is a dimensionless parameter.

Next, we use the matrix of incompatible shape functions based on quadratic interpolation, whose derivation is thoroughly described in the book by Obraztsov, Savel'ev, and Khazanov [43]

$$\alpha_r^{inc} = \begin{pmatrix} \dfrac{(1+\xi_r \xi)(1+\eta_r \eta)}{4} & -\dfrac{\xi_r \eta_r}{8}\left(\dfrac{\mu}{\gamma}\xi^2 + \gamma \eta^2 - \dfrac{\mu}{\gamma} - \gamma\right) \\ -\dfrac{\xi_r \eta_r}{8}\left(\dfrac{1}{\gamma}\xi^2 + \mu \gamma \eta^2 - \dfrac{1}{\gamma} - \mu \gamma\right) & \dfrac{(1+\xi_r \xi)(1+\eta_r \eta)}{4} \end{pmatrix}. \tag{13}$$

By substituting equation (13) into equation (2), we obtain an alternative expression for the matrix relating nodal displacements to strains

$$\beta_r^{inc} = \frac{1}{2} \begin{pmatrix} \dfrac{\xi_r(1+\eta_r\eta)}{a_{fe}} & -\dfrac{\mu\eta_r\xi_r\xi}{b_{fe}} \\ -\dfrac{\mu\xi_r\eta_r\eta}{a_{fe}} & \dfrac{\eta_r(1+\xi_r\xi)}{b_{fe}} \\ \dfrac{\eta_r}{b_{fe}} & \dfrac{\xi_r}{a_{fe}} \end{pmatrix}. \tag{14}$$

Substituting equation (14) into equations (7) and (8), we obtain analytical expressions for the submatrices representing normal (equation 15) and shear (equation 16) deformations in the stiffness matrix of the incompatible finite element for an isotropic material

$$k_{r,s}^{inc.E} = \frac{Eh}{4(1-\mu^2)} \begin{pmatrix} \gamma\xi_r\xi_s\left(1+\eta_r\eta_s\dfrac{1-\mu^2}{3}\right) & \mu\xi_r\eta_s \\ \mu\eta_r\xi_s & \dfrac{\eta_r\eta_s}{\gamma}\left(1+\xi_r\xi_s\dfrac{1-\mu^2}{3}\right) \end{pmatrix}, \tag{15}$$

$$k_{r,s}^{inc.G} = \frac{Gh}{4} \begin{pmatrix} \dfrac{\eta_r\eta_s}{\gamma} & \eta_r\xi_s \\ \xi_r\eta_s & \gamma\xi_r\xi_s \end{pmatrix}. \tag{16}$$

The condition of displacement compatibility between adjacent incompatible elements is ensured only at the nodes, while displacements along the element interfaces will experience discontinuities. Nevertheless, the use of such elements allows high accuracy to be achieved [43].

To derive a rectangular finite element for an orthotropic material, we use the specific elasticity matrix under plane stress [44, 46]

$$\chi^{ort} = \begin{pmatrix} \dfrac{1}{E_x} & -\dfrac{\mu_{xy}}{E_y} & 0 \\ -\dfrac{\mu_{xy}}{E_y} & \dfrac{1}{E_y} & 0 \\ 0 & 0 & \dfrac{1}{G_{xy}} \end{pmatrix}^{-1} = \frac{1}{1-n_{xy}\mu_{xy}^2} \begin{pmatrix} E_x & E_x\mu_{xy} & 0 \\ E_x\mu_{xy} & E_y & 0 \\ 0 & 0 & \dfrac{1-n_{xy}\mu_{xy}^2}{G_{xy}} \end{pmatrix}, \tag{17}$$

Here, $E_x$ and $E_y$ are the elastic moduli in the $x$- and $y$-direction, respectively; $n_{xy} = E_x/E_y$; $\mu_{xy}$ is the Poisson's ratio in the $xy$-plane, measured under a force applied in the $x$-direction; and $G_{xy}$ is the shear modulus in the $xy$-plane. The accuracy of this elasticity matrix is maintained as long as there is no significant difference between the Poisson's ratios $\mu_{xy}$ and $\mu_{yx}$, measured in perpendicular directions within the computational plane.

The submatrices $\chi_E^{ort}$ and $\chi_G^{ort}$ are defined by

$$\chi_E^{ort} = \frac{1}{1-n_{xy}\mu_{xy}^2} \begin{pmatrix} E_x & E_x\mu_{xy} & 0 \\ E_x\mu_{xy} & E_y & 0 \\ 0 & 0 & 0 \end{pmatrix}, \tag{18}$$

and

$$\chi_G^{ort} = G_{xy} \begin{pmatrix} 0 & 0 & 0 \\ 0 & 0 & 0 \\ 0 & 0 & 1 \end{pmatrix}. \tag{19}$$

After substituting equations (2) and (18) into equation (7) and equations (2) and (19) into equation (8), we obtain analytical expressions for the submatrices representing normal (equation 20) and shear (equation 21) deformations in the stiffness matrix of a compatible finite element using the classical shape functions for an orthotropic material

$$k_{r,s}^{o.E} = \frac{h}{4(1-n_{xy}\mu_{xy}^2)} \begin{pmatrix} E_x \gamma \xi_r \xi_s \left(1 + \frac{\eta_r \eta_s}{3}\right) & E_x \mu_{xy} \xi_r \eta_s \\ E_x \mu_{xy} \eta_r \xi_s & E_y \frac{\eta_r \eta_s}{\gamma} \left(1 + \frac{\xi_r \xi_s}{3}\right) \end{pmatrix}, \tag{20}$$

$$k_{r,s}^{o.G} = \frac{G_{xy} h}{4} \begin{pmatrix} \frac{\eta_r \eta_s}{\gamma} \left(1 + \frac{\xi_r \xi_s}{3}\right) & \eta_r \xi_s \\ \xi_r \eta_s & \gamma \xi_r \xi_s \left(1 + \frac{\eta_r \eta_s}{3}\right) \end{pmatrix}. \tag{21}$$

By substituting equations (14) (with $\mu$ replaced by $\mu_{xy}$) and (18) into equation (7) and equations (14) and (19) into equation (8) we obtain analytical expressions for the submatrices representing normal (equation 22) and shear (equation 23) deformations in the stiffness matrix of an incompatible finite element for an orthotropic material

$$k_{r,s}^{inc.o.E} = \frac{h}{4(1-n_{xy}\mu_{xy}^2)}$$
$$\times \begin{pmatrix} \gamma \xi_r \xi_s \left(E_x + \frac{\eta_r \eta_s \left(E_x - 2E_x \mu_{xy}^2 + E_y \mu_{xy}^2\right)}{3}\right) & E_x \mu_{xy} \xi_r \eta_s \\ E_x \mu_{xy} \eta_r \xi_s & \frac{\eta_r \eta_s}{\gamma} \left(E_y + \frac{\xi_r \xi_s \left(E_y - E_x \mu_{xy}^2\right)}{3}\right) \end{pmatrix}, \tag{22}$$

$$k_{r,s}^{inc.o.G} = \frac{G_{xy} h}{4} \begin{pmatrix} \frac{\eta_r \eta_s}{\gamma} & \eta_r \xi_s \\ \xi_r \eta_s & \gamma \xi_r \xi_s \end{pmatrix}. \tag{23}$$

The general technical aspects of the finite element algorithms employed in the calculations are thoroughly detailed in the previous articles [45, 47]. In the study [45], the effectiveness of these algorithms was demonstrated by the good agreement among the results obtained from the plane modeling of layered composites with honeycomb cores, three-dimensional simulations using the Comsol Multiphysics system, and laboratory experiments conducted on samples produced via additive manufacturing. Here, we primarily focus on the key expressions essential to the discussion.

## 3. Modeling the mechanical state in isotropic auxetics under shear and indentation

In this chapter, we initially analyze auxetic plates subjected to shear loading, as shown in Figure 2. The dimensions $a$, $b$, and $h$ correspond to the $x$-, $y$-, and $z$-axes, respectively. The computational domain in the $xy$-plane is divided into rectangular finite elements with varying mesh densities.

The boundary condition of a fixed connection is applied to the bottom face

$$u_{x,y,z}\left(0 \le x \le a, y = 0, 0 \le z \le h\right) = 0, \tag{24}$$

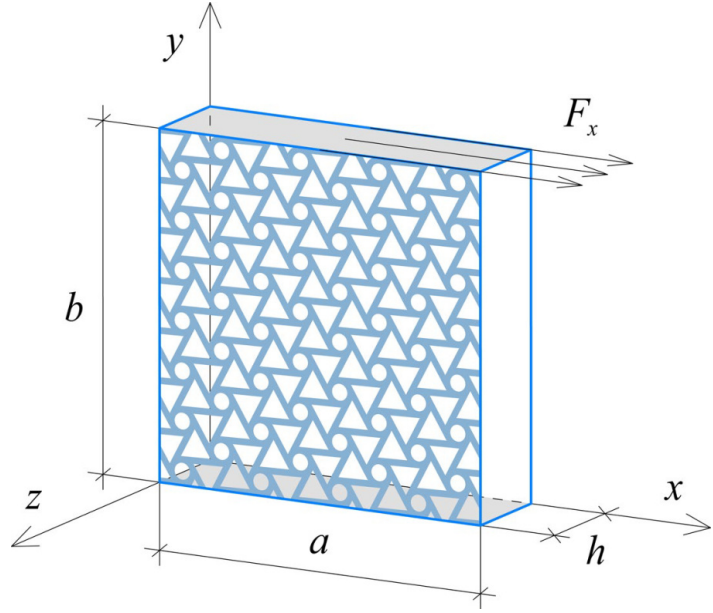

**Fig. 2.** Boundary condition, loading scheme, and dimensions of an analyzed plate under shear.

and a uniformly distributed shear force is exerted on the upper face

$$F_x = F_x\left(0 \le x \le a, y = b, 0 \le z \le h\right). \tag{25}$$

Next, we analyze the same plates under indentation loading (see Fig. 3), where the boundary condition of a fixed connection (24) remains on the bottom face, and a uniformly distributed vertical force is applied over a segment on the top face

$$F_y = F_y\left(x = a/2, y = b, 0 \le z \le h\right). \tag{26}$$

To prevent numerical singularity in the load zone, two additional equal layers of finite elements with a total thickness of $t$ are added to the upper face. These layers are not considered when calculating displacements and stresses.

Equivalent stresses are determined according to the von Mises criterion of a limit state [48, 49] relevant to the plane problem

$$\sigma_e = \sqrt{\sigma_1^2 + \sigma_2^2 - \sigma_1 \sigma_2}, \tag{27}$$

where $\sigma_1$ and $\sigma_2$ are the principal stresses

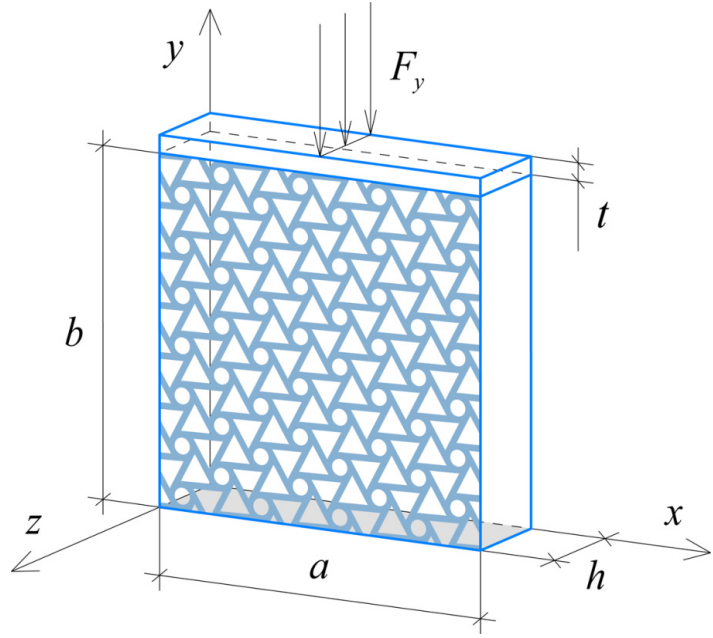

**Fig. 3.** Boundary condition, loading scheme, and dimensions of an analyzed plate under indentation.

$$\sigma_{1,2} = \frac{\sigma_x + \sigma_y}{2} \pm \sqrt{\left(\frac{\sigma_x - \sigma_y}{2}\right)^2 + \tau_{xy}^2}, \tag{28}$$

$\sigma_x$ and $\sigma_y$ are the normal stresses, and $\tau_{xy}$ is the shear stress.

The maximum shear stress is expressed by

$$\tau_{max} = \sqrt{\left(\frac{\sigma_x - \sigma_y}{2}\right)^2 + \tau_{xy}^2}. \tag{29}$$

### 3.1. Stress-strain state in isotropic auxetic materials under shear

Let us consider plates (Fig. 2) with dimensions $a = b = 15$ mm and $h = 1.5$ mm, exposed to a shear force of $F_x = 80$ N. The modulus of elasticity is assumed constant at $E = 2.8$ GPa, while the shear modulus $G = G(E, \mu)$ varies according to the well-known equation

$$G = \frac{E}{2(1+\mu)}, \tag{30}$$

as a result of changes in Poisson's ratio within the range $-1 < \mu < 0.5$. It is evident from the equation that as Poisson's ratio decreases from 0.5 and approaches –1, the shear modulus tends toward infinity. This means a significant increase in the material's resistance to shear, which in turn leads to a substantial reduction in displacements under the applied force. Figure 4 illustrates this behavior, showing the relationship between maximum displacement and Poisson's ratio for both the classical (equation 4) and incompatible (equation 13) shape functions over different mesh element densities.

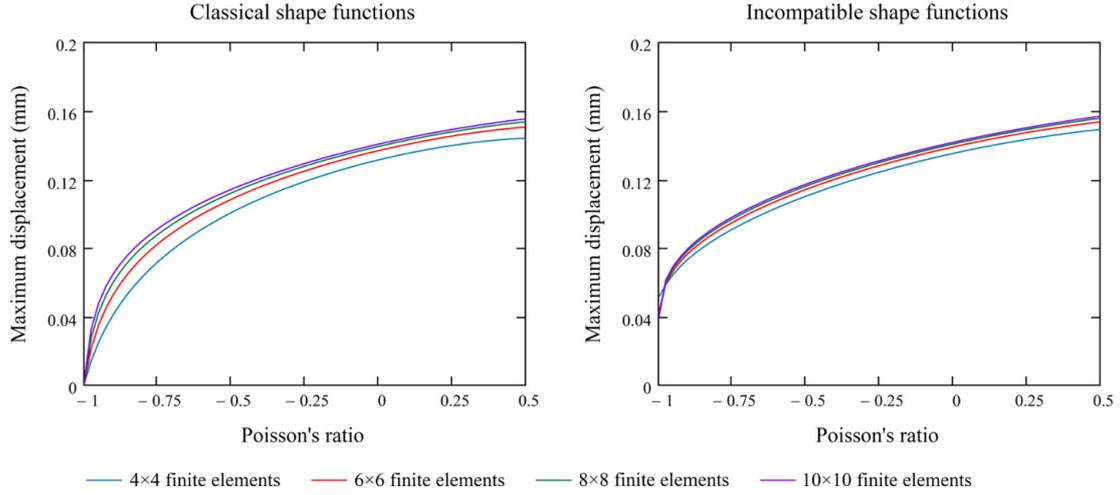

**Fig. 4.** Dependence of maximum displacement on Poisson's ratio under shear loading for the classical and incompatible shape functions across mesh element densities ranging from 4×4 to 10×10.

Despite the similar impact of the classical and incompatible shape functions in modeling displacements, they do not yield consistent results in stress modeling across the entire range of Poisson's ratio variation (Fig. 5).

According to Hooke's law in its Lame form [41-44]

$$\begin{pmatrix} \sigma_x \\ \sigma_y \\ \tau_{xy} \end{pmatrix} = \chi \begin{pmatrix} \varepsilon_x \\ \varepsilon_y \\ \gamma_{xy} \end{pmatrix}, \qquad (31)$$

the stress vector is a function of the strain vector, which is obtained through a matrix that relates nodal displacements to strains

$$\begin{pmatrix} \varepsilon_x \\ \varepsilon_y \\ \gamma_{xy} \end{pmatrix} = \beta_r \begin{pmatrix} v_x \\ v_y \end{pmatrix}, \qquad (32)$$

using shape functions. As Poisson's ratio decreases within the negative range, the components of the elastic matrix for plane stress (equation 5) undergo different qualitative changes. Specifically, the diagonal elements $\chi_{1,1} = \chi_{2,2}$, including the shear modulus $\chi_{3,3}$, increase, whereas the off-diagonal elements $\chi_{1,2} = \chi_{2,1}$ decrease. Therefore, the qualitative changes in maximum stresses depend on the magnitude of the reduction in maximum strains and their respective directions, which are determined by shape functions.

As shown in Figure 5, the use of incompatible shape functions allows for obtaining characteristic auxetic behavior over a wide range of negative Poisson's ratios. This behavior is marked by a reduction in maximum equivalent stresses due to more efficient stress distribution [3, 5-7] while maintaining equilibrium conditions

$$\begin{aligned} \frac{1}{N} \sum_{i=1}^{N} \sigma_i^x &= 0, \\ \frac{1}{N} \sum_{i=1}^{N} \sigma_i^y &= 0, \\ \frac{1}{N} \sum_{i=1}^{N} \tau_i^{xy} &= \frac{F_x}{ah}, \end{aligned} \qquad (33)$$

where $\sigma_i^x$ and $\sigma_i^y$ are the average normal stresses along the *x*- and *y*-axes within the *i*-th finite element, respectively, $\tau_i^{xy}$ denotes the average shear stress in the *i*-th element, and *N* is the total number of finite elements.

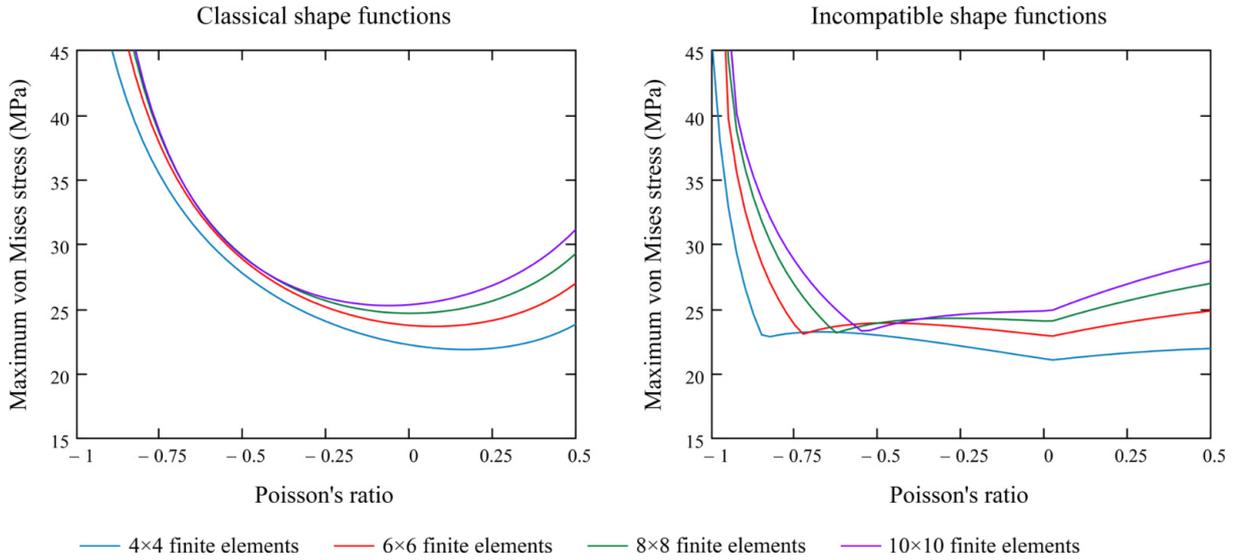

**Fig. 5.** Dependence of maximum von Mises stress on Poisson's ratio under shear loading for the classical and incompatible shape functions across mesh element densities ranging from 4×4 to 10×10.

In contrast, the classical shape functions lead to an almost immediate increase in maximum von Mises stresses as Poisson's ratio decreases below zero, which is particularly evident in Figure 6.

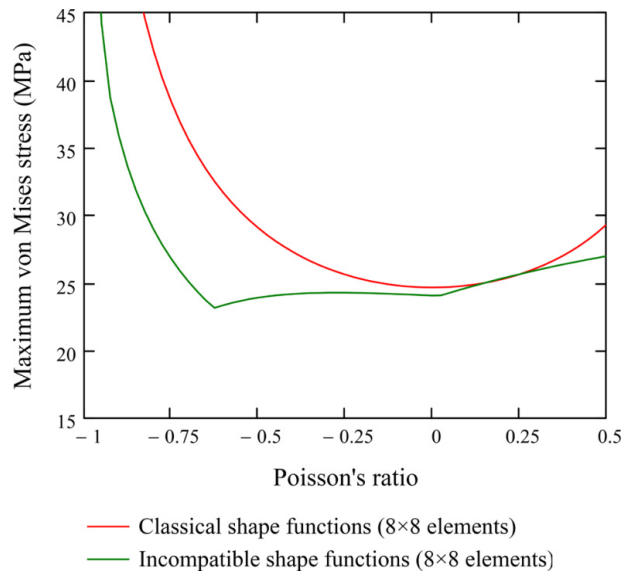

**Fig. 6.** Comparison of graphs showing the dependence of maximum von Mises stress on Poisson's ratio under shear loading for the classical and incompatible shape functions with a mesh element density of 8×8.

Along with comparing the maximum equivalent stresses obtained via different shape functions, both the maximum principal stresses (Fig. 7) and the highest maximum shear stresses (Fig. 8) demonstrate similar differences in variation. Under shear loading, the maximum principal stresses along the *x*- and *y*-axes are equal when considered in absolute terms.

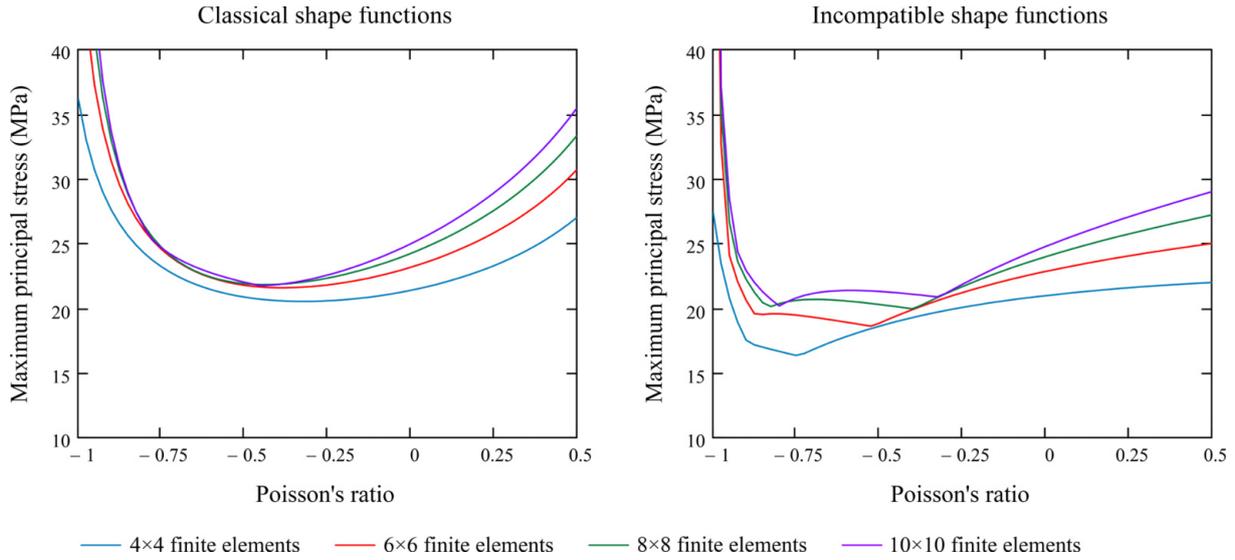

**Fig. 7.** Dependence of maximum principal stress on Poisson's ratio under shear loading for the classical and incompatible shape functions across mesh element densities ranging from 4×4 to 10×10.

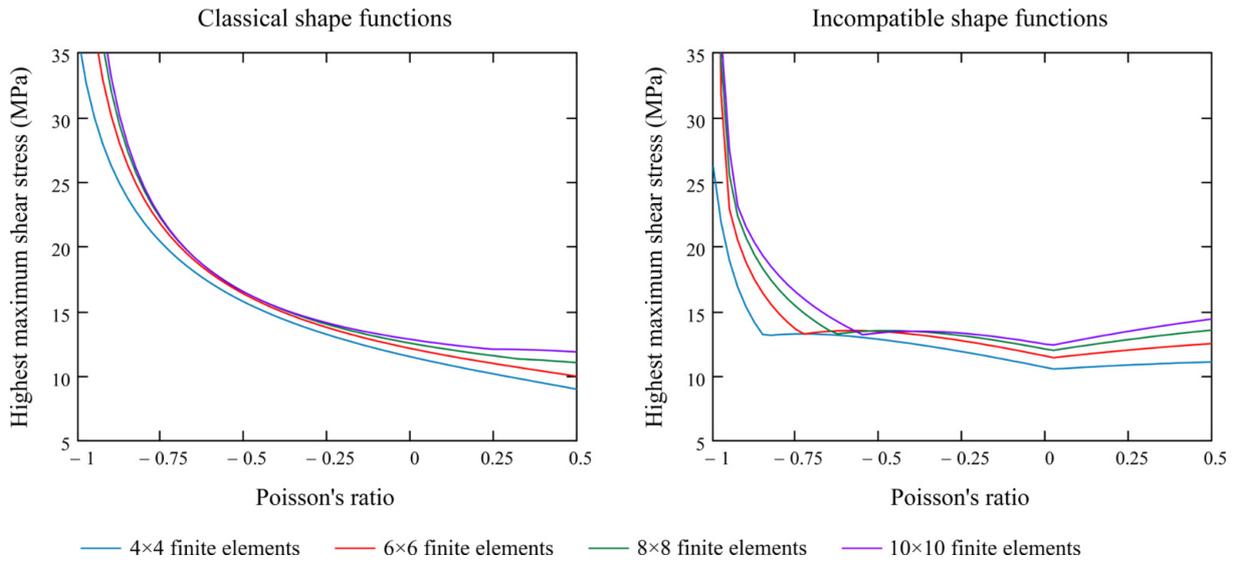

**Fig. 8.** Dependence of highest maximum shear stress on Poisson's ratio under shear loading for the classical and incompatible shape functions across mesh element densities ranging from 4×4 to 10×10.

Additionally, the classical shape functions lead to the emergence of "ripples" in von Mises stress distributions for materials with a negative Poisson's ratio, unlike the incompatible shape functions. This distinction is clearly illustrated by the contour plots for various Poisson's ratios ranging from -0.6 to -0.2 (Fig. 9). Notably, "ripples" initially emerge in maximum shear stress distributions (Fig. 10), which then propagate to equivalent stress distributions. This effect becomes more pronounced at lower Poisson's ratios, where the stress contour plots exhibit significant distortion. On the other hand, for positive and zero Poisson's ratios, "ripples" are weakly expressed (Figs. 11 and 12).

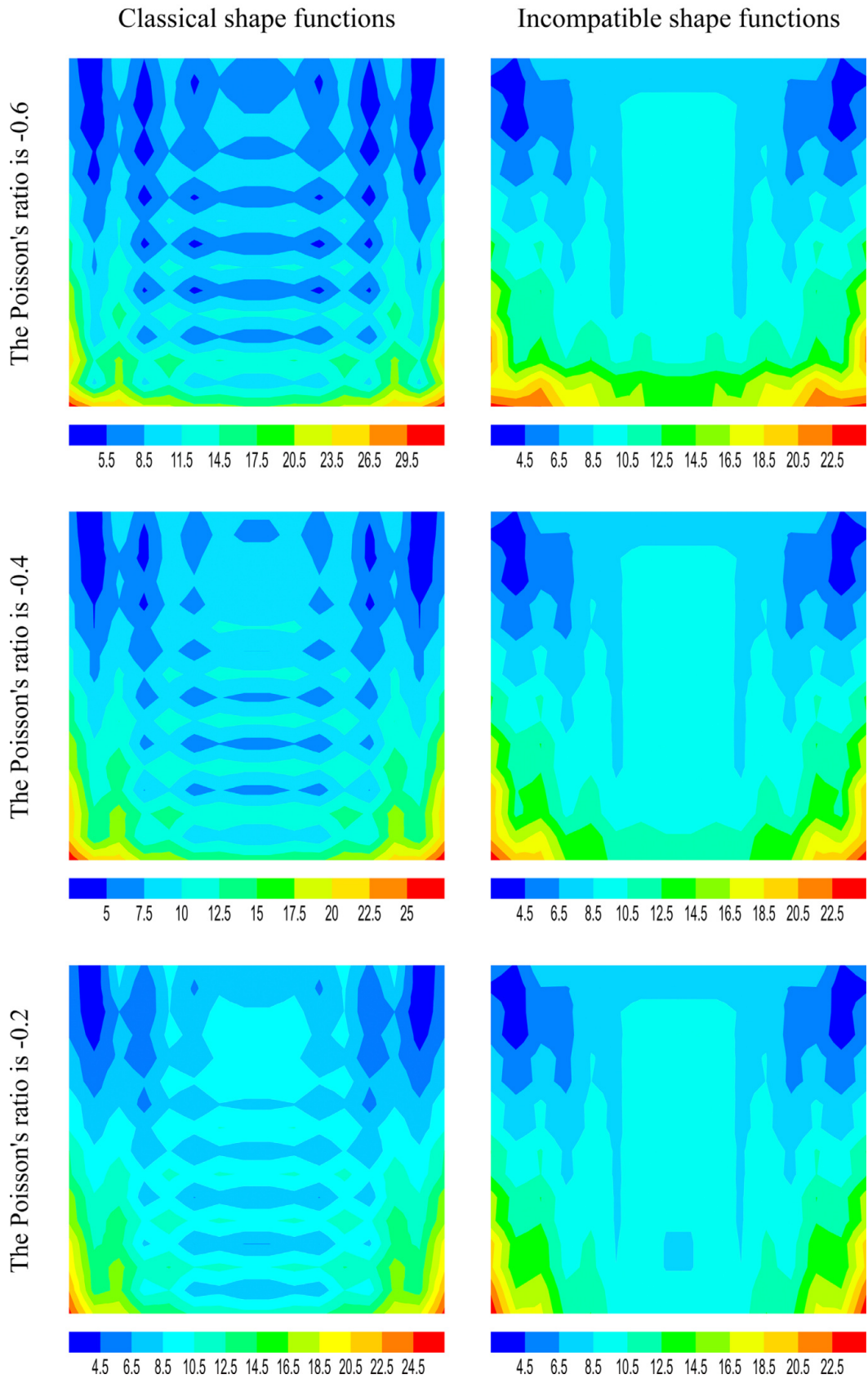

**Fig. 9.** Contour plots illustrating the von Mises stress distribution (in MPa) for different negative Poisson's ratios in an isotropic material, derived using the classical and incompatible shape functions.

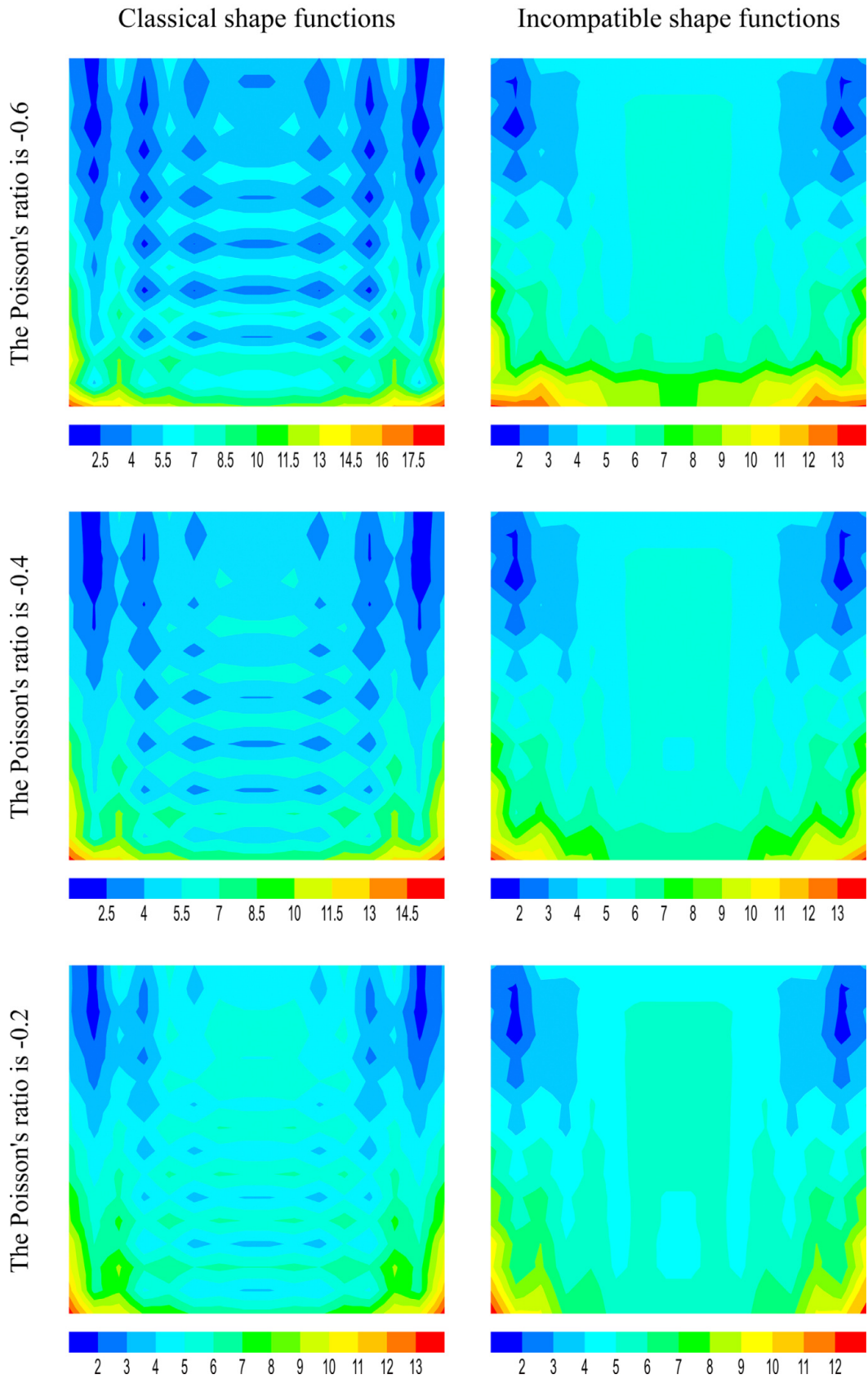

**Fig. 10.** Contour plots illustrating the maximum shear stress distribution (in MPa) for different negative Poisson's ratios in an isotropic material, derived using the classical and incompatible shape functions.

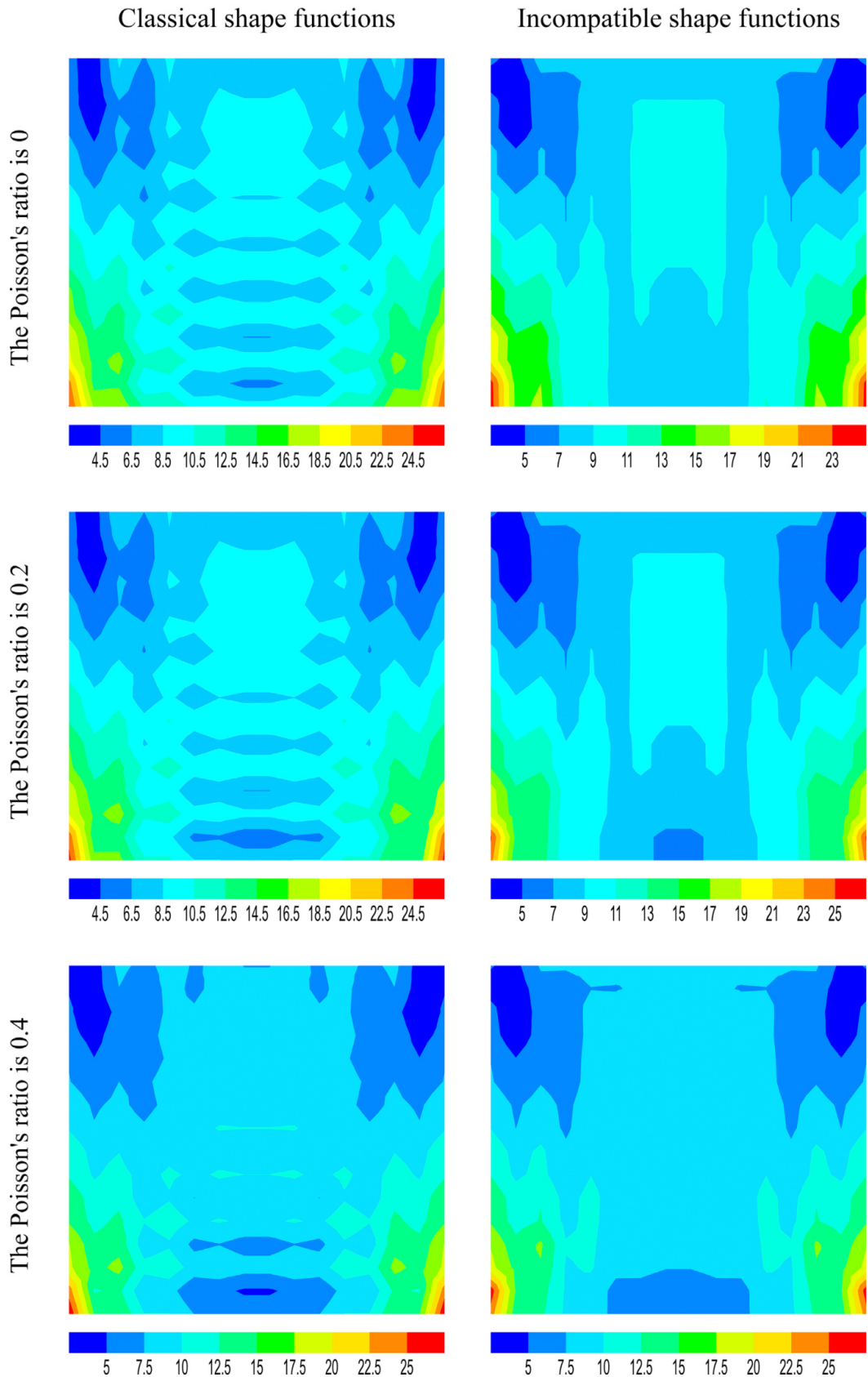

**Fig. 11.** Contour plots illustrating the von Mises stress distribution (in MPa) for different positive and zero Poisson's ratios in an isotropic material, derived using the classical and incompatible shape functions.

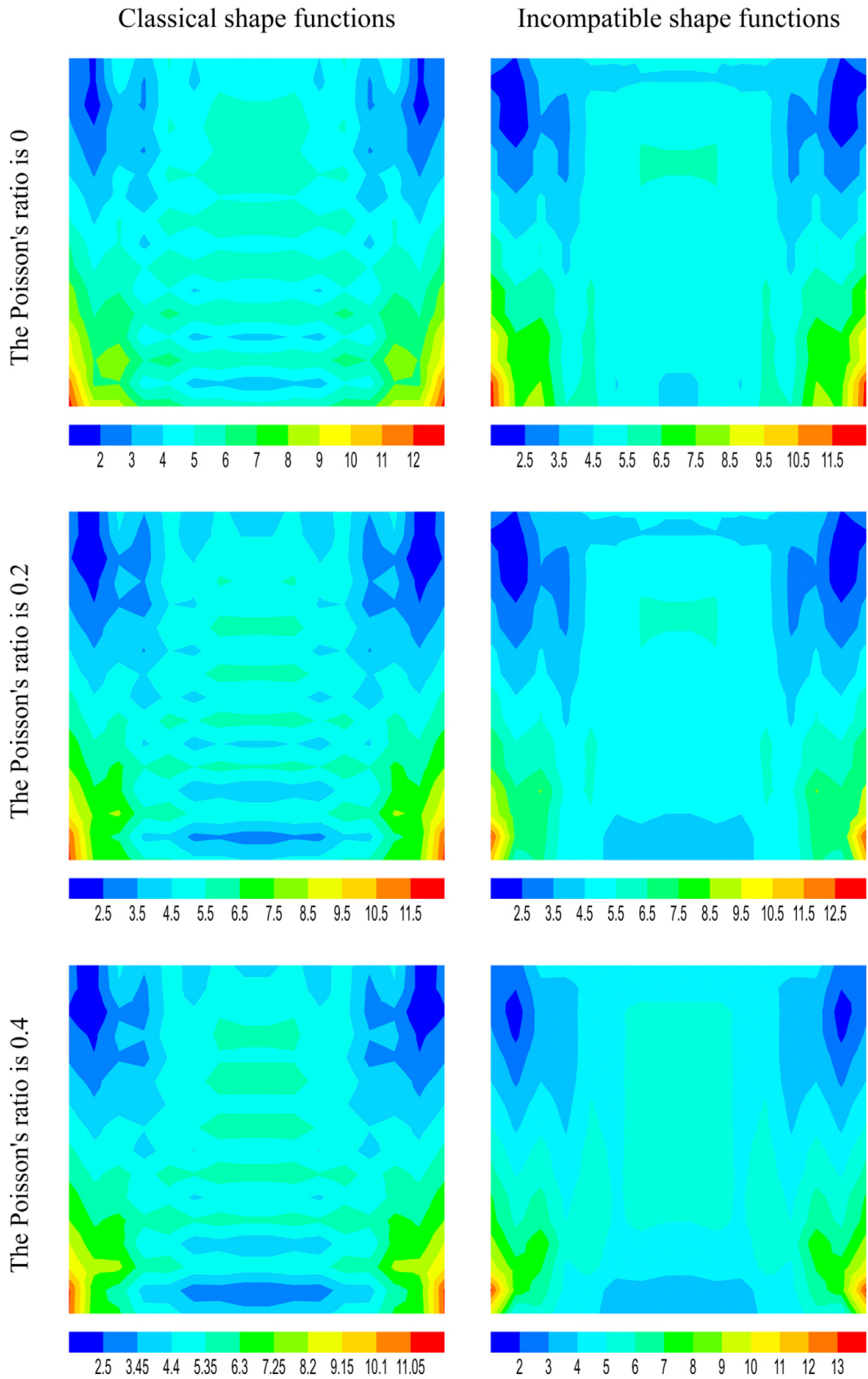

**Fig. 12.** Contour plots illustrating the maximum shear stress distribution (in MPa) for different positive and zero Poisson's ratios in an isotropic material, derived using the classical and incompatible shape functions.

## 3.2. Stress-strain state in isotropic auxetic materials under indentation

In this section, we consider plates with the geometric and material parameters given above, subjected to an indentation force of $F_y = 120$ N (see Fig. 3). The total thickness of the additional layers is $t = 2$ mm.

The classical and incompatible shape functions yield qualitatively similar results in displacement modeling (Fig. 13) compared to those obtained under the shear conditions. Stress modeling also demonstrates similar differences (Fig. 14) relative to the shear conditions, which favorably distinguish the incompatible shape functions in their ability to describe characteristic auxetic behavior (Fig. 15).

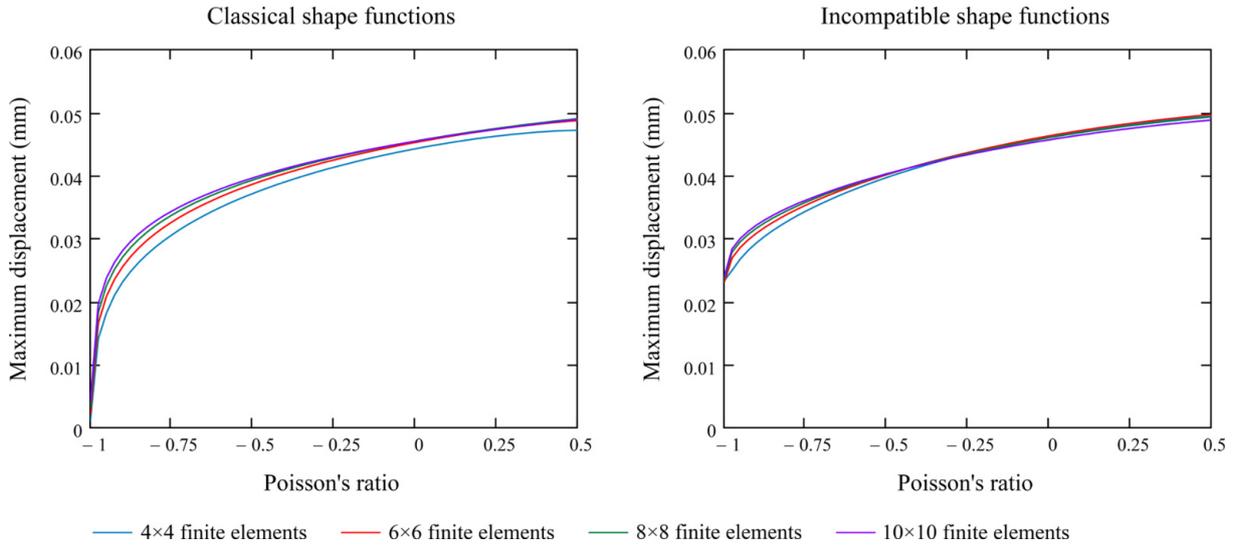

**Fig. 13.** Dependence of maximum displacement on Poisson's ratio under indentation loading for the classical and incompatible shape functions across mesh element densities ranging from 4×4 to 10×10.

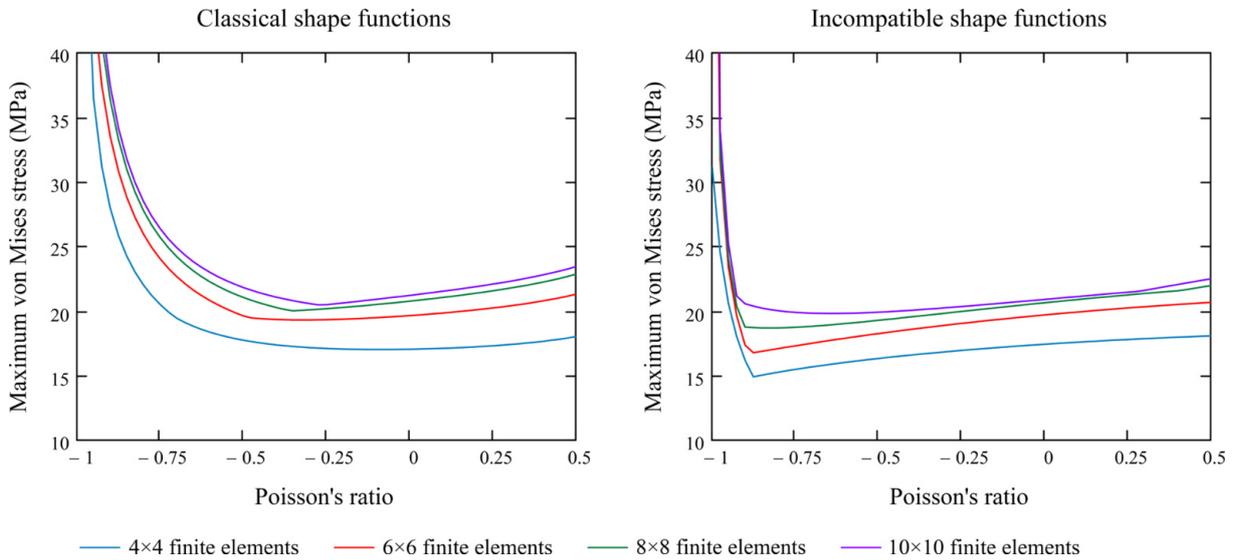

**Fig. 14.** Dependence of maximum von Mises stress on Poisson's ratio under indentation loading for the classical and incompatible shape functions across mesh element densities ranging from 4×4 to 10×10.

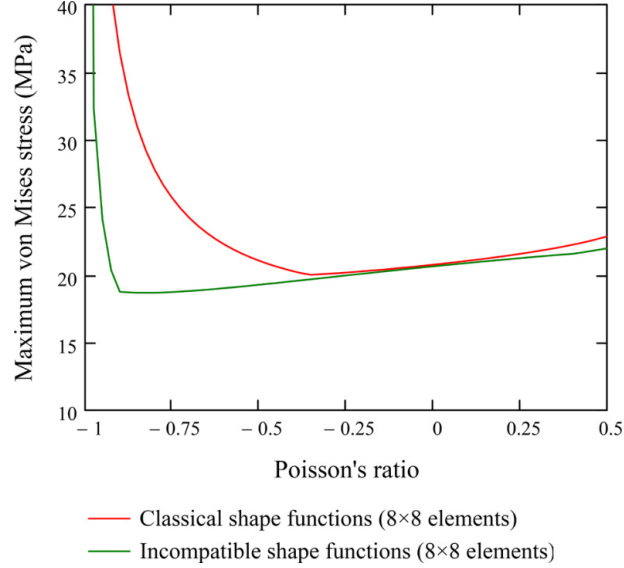

**Fig. 15.** Comparison of graphs showing the dependence of maximum von Mises stress on Poisson's ratio under indentation loading for the classical and incompatible shape functions with a mesh element density of 8×8.

Thus, unlike the incompatible shape functions (equation 13), the classical shape functions (equation 4) are considerably less effective in providing qualitatively accurate results when modeling the mechanical behavior of isotropic auxetics.

**4. Modeling the mechanical state in auxetic honeycombs under shear and indentation**

Let us now examine hexachiral honeycomb plates. This auxetic structure (Fig. 16) possesses a sixfold, rotationally symmetric lattice composed of cylindrical nodes interconnected by tangentially attached ligaments, thereby forming a hexagonal pattern that lacks mirror symmetry [2-4, 6, 8, 34, 39].

The hexachiral honeycombs are defined by the following geometric parameters (see Fig. 16): $L_h$ is the length of the tangentially attached ligaments, $r_h$ is the medium radius of the cylinders, $R_h$ is the distance between the centers of two cylinders connected by a common ligament, $\theta_h$ is the angle between the tangentially attached ligament and the line connecting the centers of the cylinders, and $t_h$ is the wall thickness of the honeycomb structure.

The in-plane effective modulus of elasticity for hexachiral honeycombs is given by the following expression [34]

$$E_x^{hex} = E_y^{hex} = E_s \frac{4\sqrt{3}\left((t_h/L_h)^3 + (t_h/L_h)^5\right)}{2(t_h/L_h)^4 \cos^2\theta_h + 2\sin^2\theta_h + 6(t_h/L_h)^2}, \qquad (34)$$

where $\cos\theta_h = L_h/R_h$, $\sin\theta_h = \sqrt{R_h^2 - L_h^2}/R_h$, $E_s$ is the elastic modulus of the solid material constituting the honeycombs.

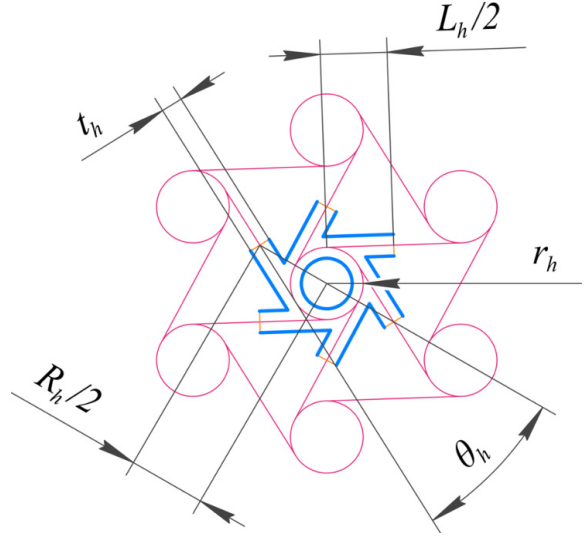

**Fig. 16.** Geometric parameters of a hexachiral honeycomb unit cell.

The effective Poisson's ratio is described by [34]

$$\mu_{xy}^{hex} = \mu_{yx}^{hex} = \frac{4(t_h/L_h)^2}{(t_h/L_h)^4 \cos^2\theta_h + 1 - \cos^2\theta_h + 3(t_h/L_h)^2} - 1. \qquad (35)$$

The effective shear modulus is determined by [34]

$$G_{xy}^{hex} = E_s\left(\frac{\sqrt{3}}{4}\left(\frac{t_h}{L_h}\right) + \frac{\sqrt{3}}{4}\left(\frac{t_h}{L_h}\right)^3\right). \qquad (36)$$

Lastly, the relative density is given by [39]

$$\rho_{rel}^{hex} = \frac{2t_h(3\cos\theta_h + \pi\sin\theta_h)}{R_h\sqrt{3}}. \qquad (37)$$

In the investigation, the following constants were accepted: $E_s = 2.8$ GPa, $L_h = 15$ mm, $r_h = 4$ mm, $R_h = 17$ mm, while $t_h$ varied within the range $1 \leq t_h \leq 3$ mm. Consequently, the graphs in Fig. 17 illustrate the variation ranges of effective material properties as $t_h$ changes, with all constants increasing as $t_h$ increases.

In addition to the hexachiral structure, we consider re-entrant honeycombs (Fig. 18). This auxetic configuration consists of a network of V-shaped unit cells with inwardly angled walls converging at sharp vertices, thereby forming a repeating bow-tie-shaped pattern that creates an interlocking framework [1-4, 6-8, 33, 39].

The re-entrant honeycombs are defined by the following geometric parameters (see Fig. 18): $B_r$ represents the horizontal size of the unit cell, $H_r$ denotes the vertical size, $h_r$ is the length of the vertical wall, $L_r$ is the length of the inclined wall, $\theta_r$ is the angle between the vertical and inclined walls, and $t_r$ is the thickness of the cell walls.

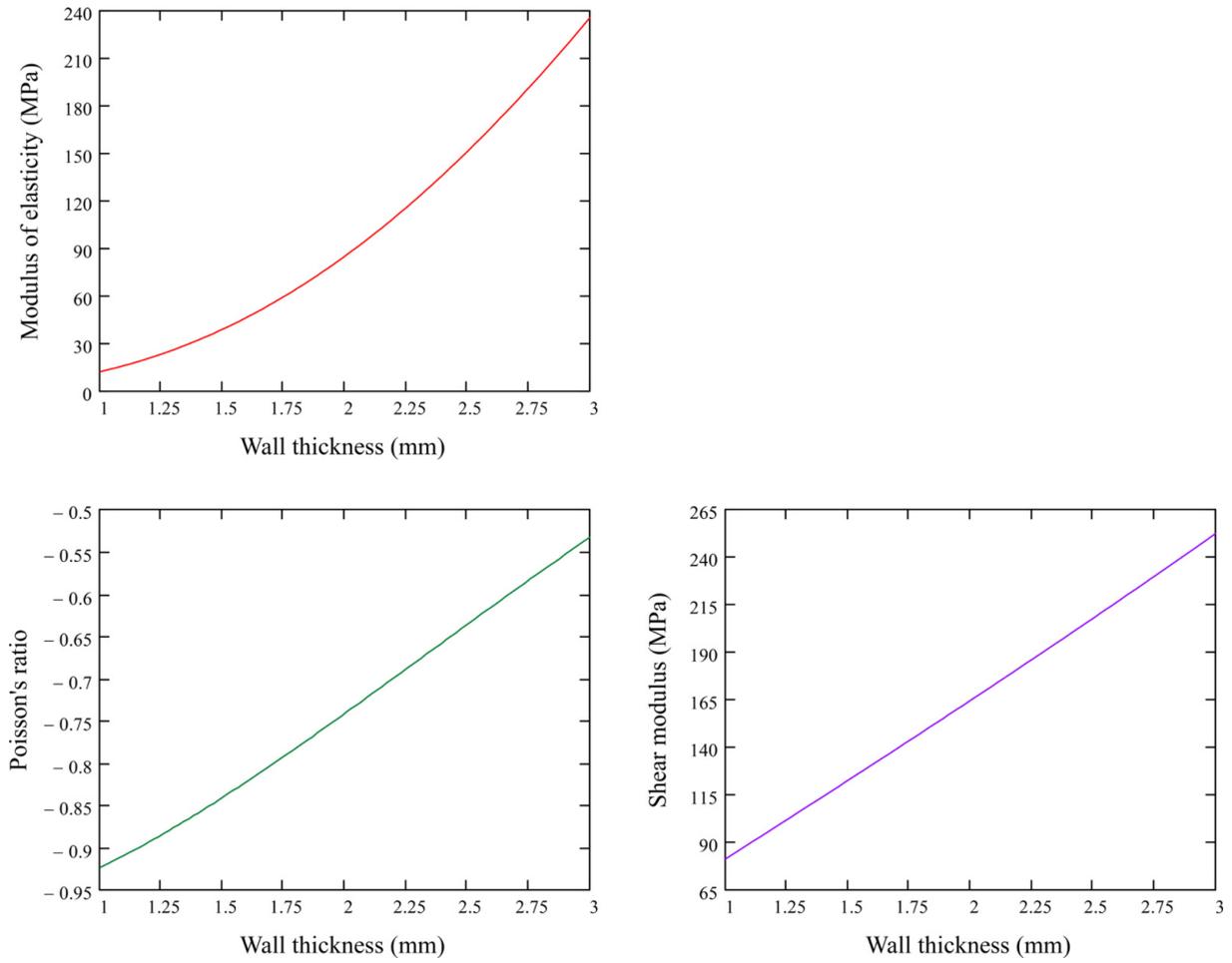

**Fig. 17.** Variation ranges of the effective constants of the hexachiral honeycombs used in the investigation.

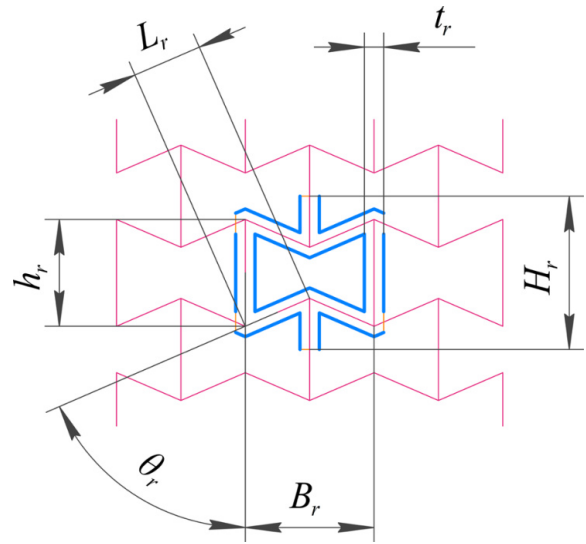

**Fig. 18.** Geometric parameters of a re-entrant honeycomb unit cell.

The in-plane effective moduli of elasticity for re-entrant honeycombs along the *x*- and *y*-axes are represented by the following expressions, respectively [33, 39]

$$E_x^{re} = E_s \frac{8t_r^3 \sin\theta_r}{B_r^2 H_r \cot^2\theta_r}, \qquad (38)$$

where $\theta_r = \arcsin(B_r/2L_r)$,

$$E_y^{re} = E_s \frac{8t_r^3 H_r \sin\theta_r}{B_r^4}. \qquad (39)$$

The effective Poisson's ratio is given by [33, 39]

$$\mu_{xy}^{re} = -\frac{B_r \cot\theta_r}{H_r}. \qquad (40)$$

The effective shear modulus is described by [33, 39]

$$G_{xy}^{re} = E_s \frac{8t_r^3 H_r \sin^3\theta_r}{B_r \left(H_r \sin\theta_r - B_r \cos\theta_r\right)^2 \left(B_r + 2H_r \sin\theta_r - 2B_r \cos\theta_r\right)}. \qquad (41)$$

Finally, the relative density is expressed by [39]

$$\rho_{rel}^{re} = \frac{t_r(H_r \sin\theta_r + B_r \cos\theta_r + 2B_r)}{B_r H_r \sin\theta_r}. \qquad (42)$$

The following constants were accepted in the investigation: $B_r = 10$ mm, $H_r = 12$ mm, and $t_r = 1.5$ mm, while $L_r$ varied within the range $5.1 \leq L_r \leq 5.9$ mm. Accordingly, the graphs in Fig. 19 demonstrate how variations in $L_r$ affect the effective material properties. As $L_r$ increases, the shear modulus rises, whereas the other constants show decreasing trends.

**4.1. Stress-strain state in auxetic honeycombs under shear**

In this section, we investigate both types of honeycomb structures, modeled as plates with dimensions $a = b = 60$ mm and $h = 6$ mm (Fig. 2), under a shear force of $F_x = 40$ N. For the calculations, an 8×8 element mesh was employed, as it was found to be sufficient to achieve reliable convergence of the numerical results (Figs. 4 and 5).

Figures 20 and 21 present graphs illustrating the dependence of maximum displacement on wall thickness $(t_h)$ for hexachiral honeycombs and on inclined wall length $(L_r)$ for re-entrant honeycombs, respectively. In hexachiral honeycombs, as the wall thickness increases from 1 to 3 mm, the Poisson's ratio increases from -0.924 to -0.533, as shown in Figure 17. Conversely, for re-entrant honeycombs, as the inclined wall length increases from 5.1 to 5.9 mm, the Poisson's ratio decreases from -0.167 to -0.522, as depicted in Figure 19.

Similar to the isotropic case in Figure 4, Figures 20 and 21 indicate no significant differences in the maximum displacement values obtained using finite elements based on the classical (equations 20 and 21) and incompatible (equations 22 and 23) shape functions.

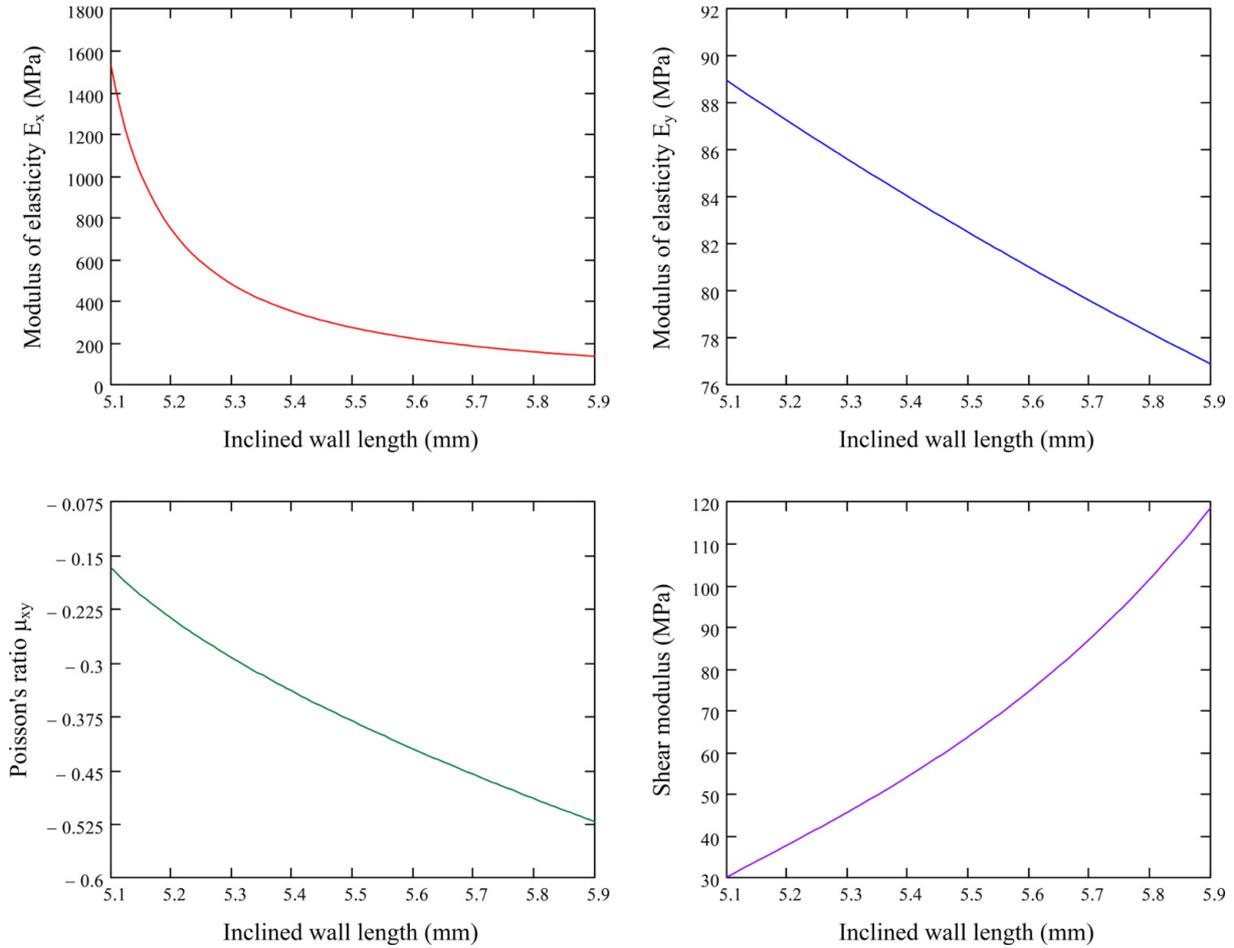

**Fig. 19.** Variation ranges of the effective constants of the re-entrant honeycombs used in the investigation.

Figures 22 and 23 illustrate the dependence of maximum von Mises stress on wall thickness for hexachiral honeycombs and on inclined wall length for re-entrant honeycombs, respectively. Across the parameter intervals, there are substantial gaps between the maximum von Mises stress values obtained using finite elements based on the classical and incompatible shape functions.

As demonstrated in the analysis of the mechanical behavior of isotropic auxetics, the incompatible shape functions are capable of providing qualitatively accurate results in modeling maximum von Mises stress changes. Therefore, it can be assumed that quantitative results from the rectangular finite element for orthotropic materials, based on the incompatible shape functions, more closely approach true values when modeling maximum equivalent stress changes compared to those obtained using the classical shape functions.

Analogous to the isotropic case, the classical shape functions result in the formation of "ripples" in stress distributions within orthotropic materials, as shown in the contour plots for hexachiral (quasi-orthotropic, Fig. 24) and re-entrant (Fig. 25) honeycombs with varying effective constants. The "ripples" also become more pronounced at lower Poisson's ratios, especially in the case of hexachiral honeycombs.

Therefore, the incompatible rectangular finite elements are appropriate for various mechanical modeling applications across positive and negative values of Poisson's ratio.

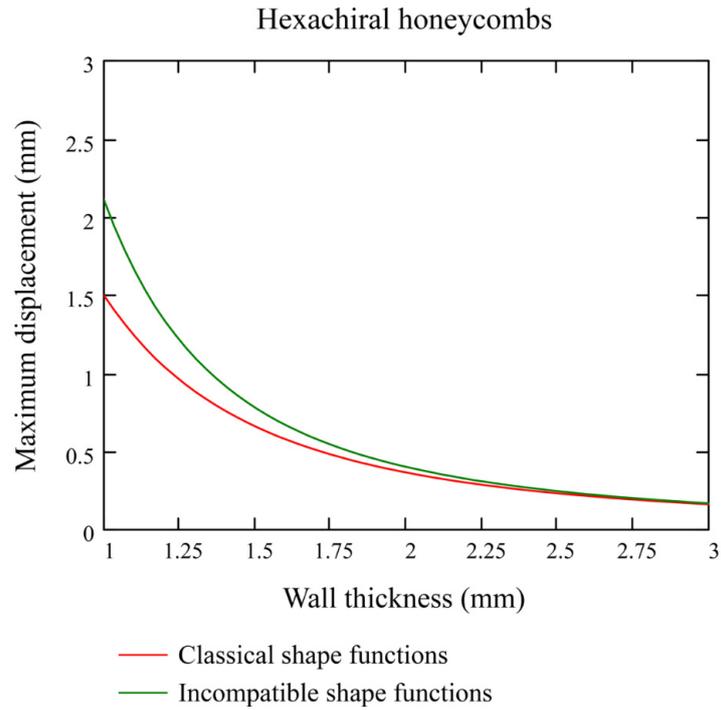

**Fig. 20.** Dependence of maximum displacement on wall thickness under shear loading for the hexachiral honeycombs.

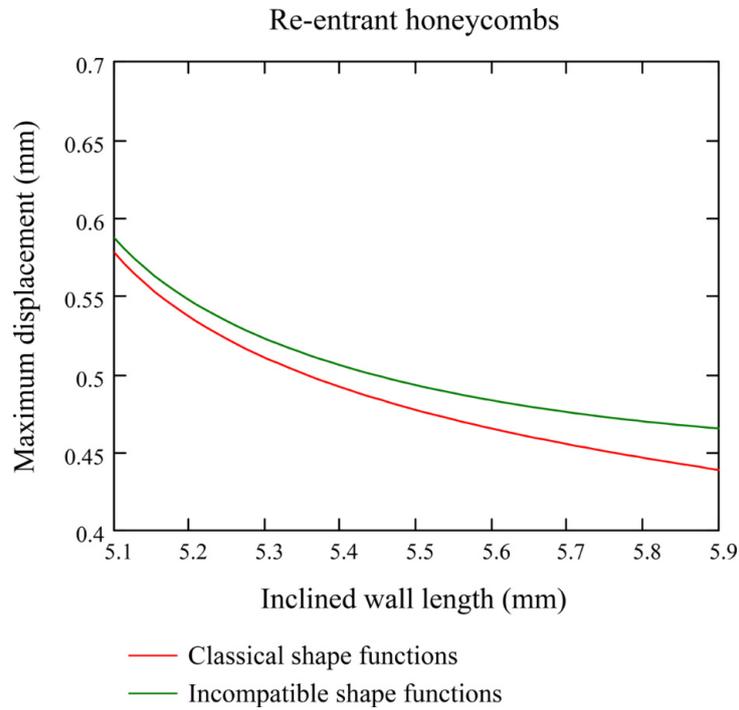

**Fig. 21.** Dependence of maximum displacement on inclined wall length under shear loading for the re-entrant honeycombs.

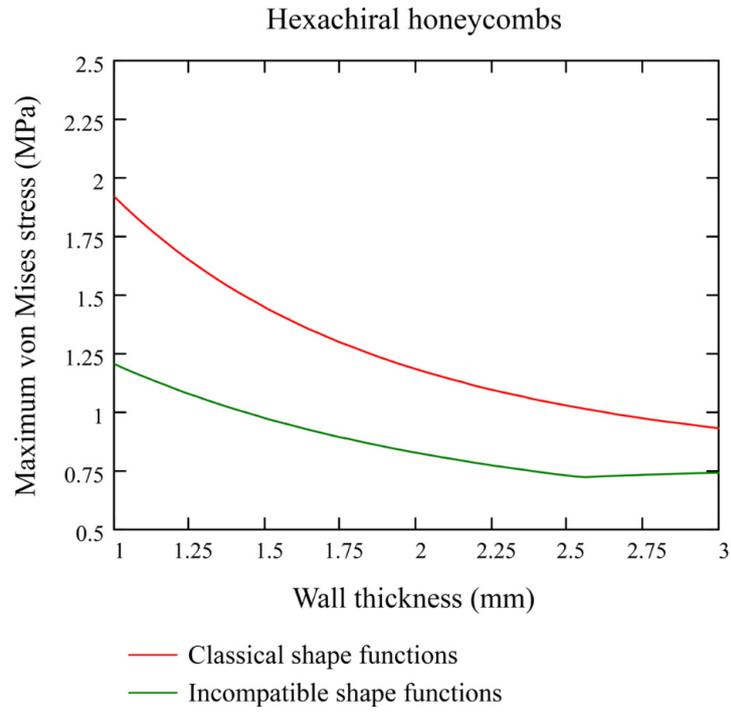

**Fig. 22.** Dependence of maximum von Mises stress on wall thickness under shear loading for the hexachiral honeycombs.

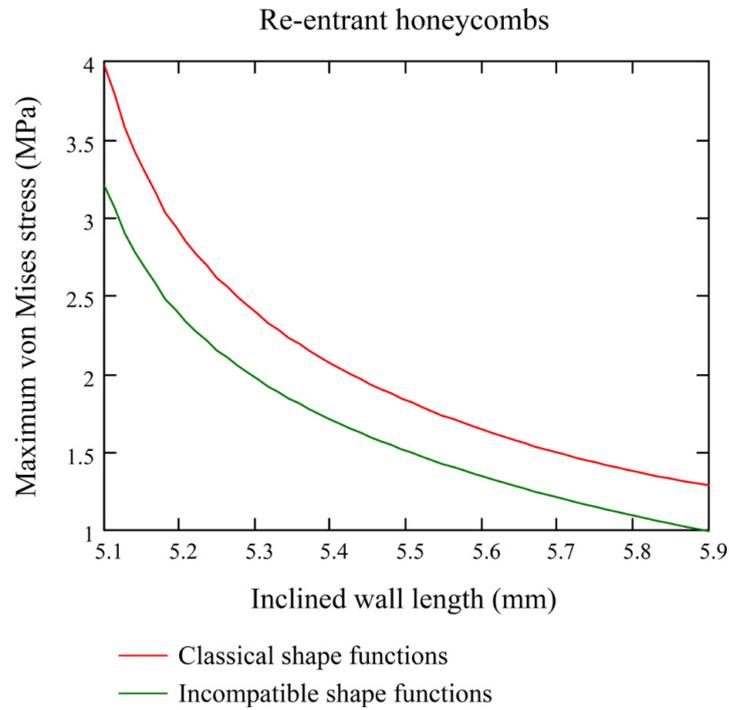

**Fig. 23.** Dependence of maximum von Mises stress on inclined wall length under shear loading for the re-entrant honeycombs.

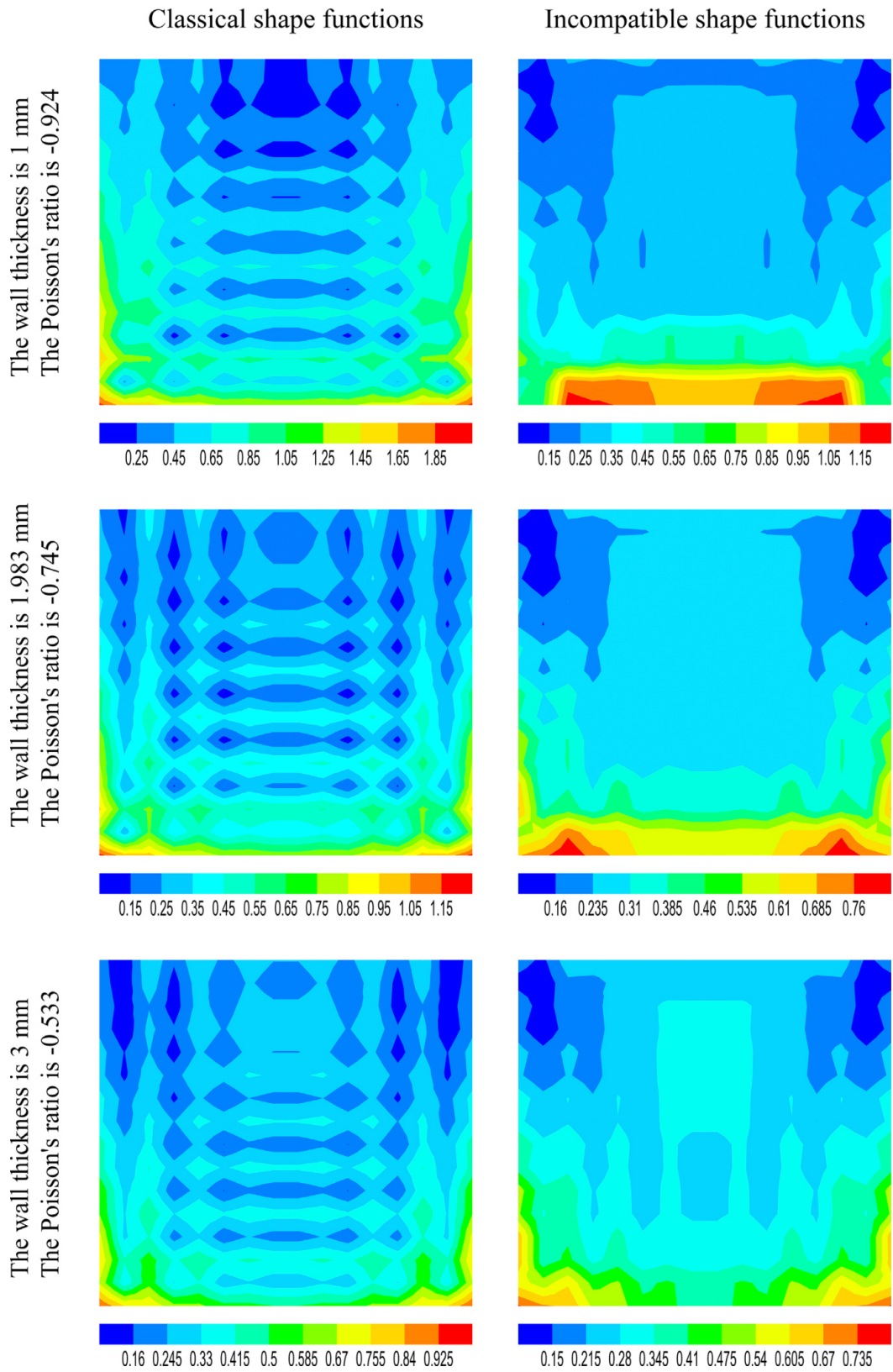

**Fig. 24.** Contour plots illustrating the von Mises stress distribution (in MPa) within the hexachiral honeycombs for different effective constants, derived using the classical and incompatible shape functions.

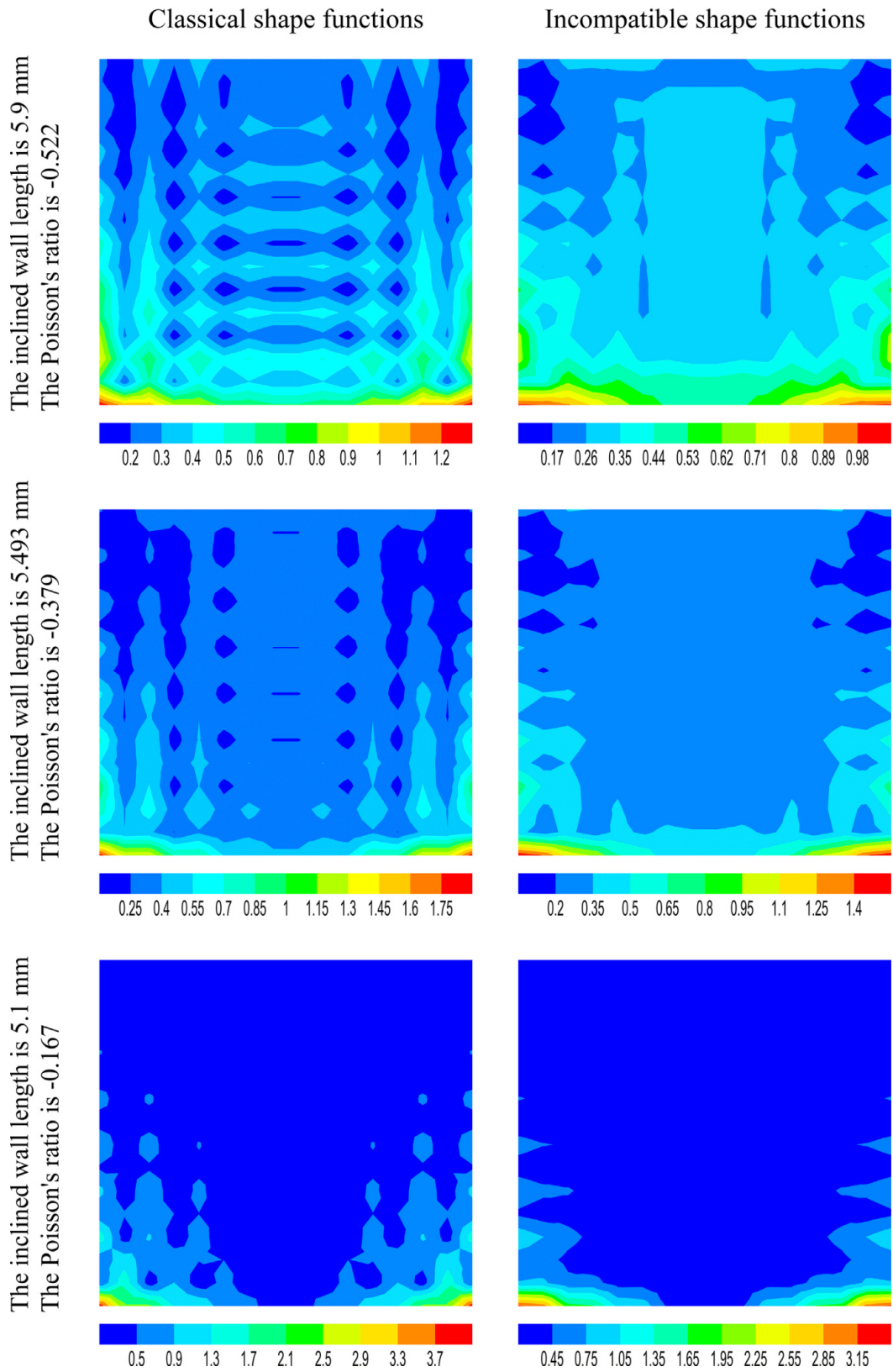

**Fig. 25.** Contour plots illustrating the von Mises stress distribution (in MPa) within the re-entrant honeycombs for different effective constants, derived using the classical and incompatible shape functions.

## 4.2. Stress-strain state in auxetic honeycombs under indentation

Next, we analyze plates undergoing an indentation force of $F_y = 60$ N (see Fig. 3), using the geometric and material parameters described in the previous section. The total thickness of the additional layers is similar to the isotropic case, which is $t = 2$ mm.

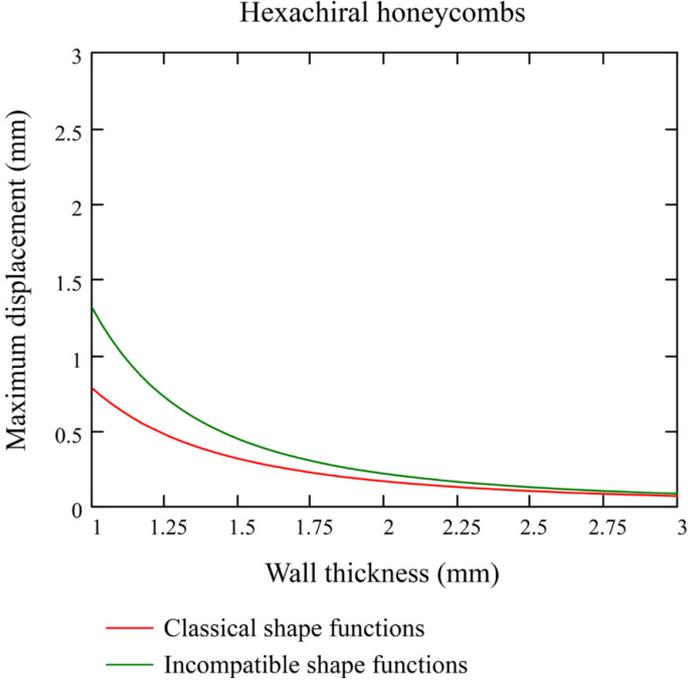

**Fig. 26.** Dependence of maximum displacement on wall thickness under indentation loading for the hexachiral honeycombs.

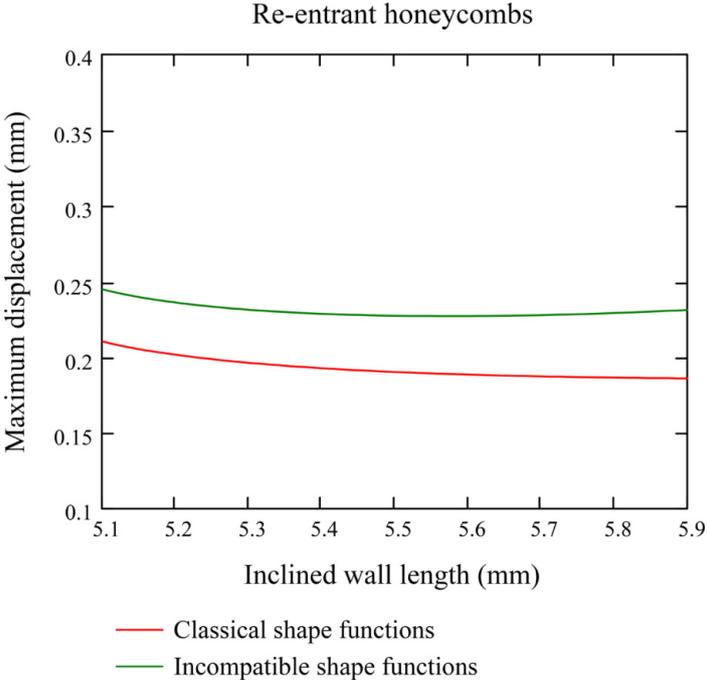

**Fig. 27.** Dependence of maximum displacement on inclined wall length under indentation loading for the re-entrant honeycombs.

Figures 26 and 27 show acceptable differences in the maximum displacement values obtained through finite elements based on the classical and incompatible shape functions. However, Figures 28 and 29 reveal substantial disparities in the maximum von Mises stresses across broad ranges of the parameter intervals. These differences result from large variations in the maximum stresses calculated using the classical shape functions.

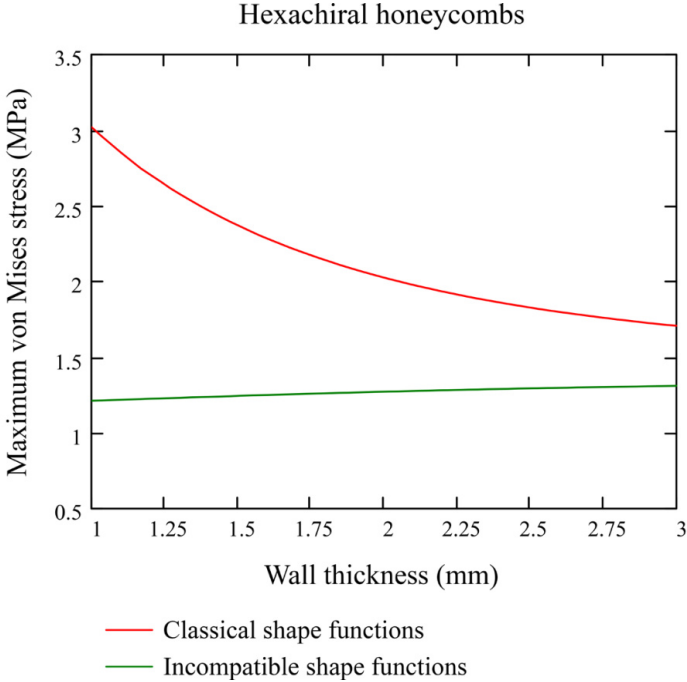

**Fig. 28.** Dependence of maximum von Mises stress on wall thickness under indentation loading for the hexachiral honeycombs.

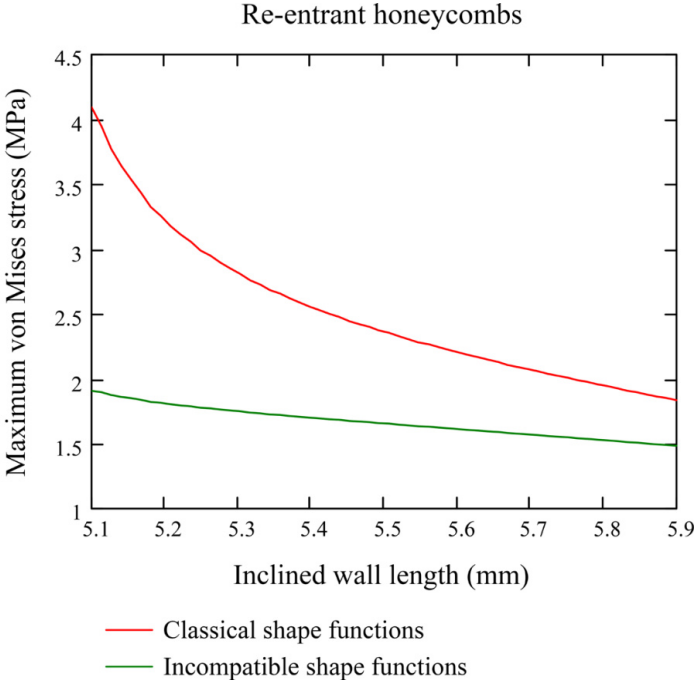

**Fig. 29.** Dependence of maximum von Mises stress on inclined wall length under indentation loading for the re-entrant honeycombs.

## 4. Conclusion

The analytical expressions of rectangular finite elements in both isotropic and orthotropic formulations for plane stress conditions are investigated in terms of their effectiveness in modeling the mechanical behavior of auxetic materials that possess a negative Poisson's ratio. Furthermore, the analytical expression of an incompatible rectangular finite element was adapted to accommodate orthotropic formulations. Both the compatible linear (classical) and incompatible quadratic shape functions were employed in the stiffness matrix formulations to model the stress-strain state of auxetic materials under shear and indentation within the linear theory of elasticity.

The findings show that while the classical and incompatible shape functions give comparable results for displacement modeling in isotropic and orthotropic materials, there are significant discrepancies in stress modeling. For isotropic auxetics, the classical shape functions are ineffective in reflecting the characteristic auxetic behavior, which involves a reduction in maximum stresses due to a more efficient stress distribution. In contrast, the incompatible shape functions are capable of providing qualitatively accurate results over a wide range of negative Poisson's ratios.

Additionally, the classical shape functions result in the emergence of "ripples" in stress distributions, leading to distorted and unreliable outcomes in isotropic auxetics and orthotropic materials with auxetic qualities, such as hexachiral (quasi-orthotropic) and re-entrant honeycombs. Conversely, stress distributions obtained using the incompatible shape functions closely align with physical responses by showing a more natural stress variation.

Therefore, the incompatible rectangular finite elements are recommended for modeling the mechanical behavior of auxetic materials. Their approximation increases the accuracy of stress calculations by correctly reflecting the qualitative behavior of the maximum stresses and the overall stress distribution. These features are important for reliable mechanical modeling of advanced materials, including engineered honeycomb structures with a negative Poisson's ratio and associated multilayer composites, where accurate stress calculations are essential for design and optimization.

Further investigations could expand the application of the incompatible rectangular finite element for an orthotropic case to a wider range of auxetic materials and their composites, as well as integrate dynamic formulations. In addition, the experimental identification of optimal criteria for the calculation of equivalent stresses in auxetic materials is essential for this area of research.